\documentclass[12pt]{article}
\usepackage{amsxtra,amssymb,amsthm,amsmath,latexsym}

\theoremstyle{plain}

\def\R{{\mathbb R}}

\def\oH{\buildrel\circ\over H}
\def\oH1{\buildrel\circ\over H\kern-.02in{}^1}
\def\oH1{\buildrel\circ\over H\kern-.02in{}^1}

\def\d{\delta}
\def\a{\alpha}
\def\a'{\alpha'}
\def\b{\beta}
\def\g{\mathcal{G}}
\def\bff{{\bf f}}
\def\bg{{\bf g}}

\usepackage{graphics}

\begin{document}

\title{
 Analysis of a method for identification of obstacles
}
\author{
A.G. Ramm\\
Mathematics Department, Kansas State University, \\
 Manhattan, KS 66506-2602, USA\\
ramm@math.ksu.edu\\ 
and\\
Semion Gutman\\
 Department of Mathematics,  University of Oklahoma, \\
 Norman, OK 73019, USA\\
sgutman@ou.edu\\ 
}

\date{}
\maketitle\thispagestyle{empty}

\centerline{ Abstract}

Some difficulties are pointed out in the methods for identification of 
obstacles based on the numerical verification of the inclusion of
a function in the range of an operator.
Numerical examples are given to illustrate theoretical conclusions.
Alternative methods of identification of obstacles are mentioned:
the Support Function Method (SFM) and the Modified Rayleigh Conjecture 
(MRC) method.



\section{Analysis }
During the last decade there are many papers published, in which
methods for identification of an obstacle are proposed, which
are based on a numerical verification of the inclusion of some function 
$f:=f(\alpha,z), \, z\in \R^3, \, \alpha \in S^2,$ in 
the range $R(B)$ of a certain operator $B$. Examples of such methods 
include  \cite{ck}, \cite{cc},\cite{Kir98}.
It is proved in this paper that the methods, proposed in the 
above papers, have essential difficulties. 
This also is demonstrated by numerical experiments.
Although it is true that
$f\not\in R(B)$ when $z\not\in D$, it turns out that in any 
neighborhood of $f$, however small, there are elements from $R(B)$. Also, 
although
$f\in R(B)$ when $z\in D$, there are elements in every neighborhood of 
$f$, however small, which do not belong to $R(B)$ even if $z\in D$. 
Therefore it is not possible to construct a
stable numerical method for identification of $D$ based on checking the 
inclusions $f\not\in R(B)$
and $f\in R(B)$. 

We prove below that the range $R(B) $ is dense in the space
$L^2(S^2)$. 

{\bf Assumption} (A): {\it We assume throughout that $k^2$ is not a 
Dirichlet eigenvalue of
the Laplacian in $D$.}

Let us introduce some {\it notations}: $N(B)$ and $R(B)$
are, respectively, the null-space and the range of a linear operator 
$B$, $D\in \R^3$ is a 
bounded 
domain (obstacle) with a smooth boundary $S$, $D'=\R^3 \setminus D,$
$u_0=e^{ik\alpha \cdot x}$, $k=const>0$, $\alpha\in S^2$ is a unit vector,
$N$ is the unit normal to $S$ pointing into $D'$, $g=g(x,y,k):=g(|x-y|):=
\frac {e^{ik|x-y|}}{4 \pi |x-y|}$, $f:=e^{-ik \alpha' \cdot z}$, where 
$z\in 
\R^3$ and $\a' \in S^2$, $\a':=xr^{-1}$,
$r=|x|$, $u=u(x,\alpha, k)$ is the scattering solution: 
$$
(\Delta +k^2)u=0 \quad in \quad D', u|_{S}=0,
 \eqno{(1)}$$ 
$$
u=u_0+v,\quad v=A(\a',\alpha,k)e^{ikr}r^{-1} +o(r^{-1}), \quad as \quad 
r\to 
\infty, \quad xr^{-1}=\a',
 \eqno{(2)}$$
where $A:=A(\a',\alpha,k)$ is called the scattering amplitude, 
corresponding 
to the obstacle $D$ and the Dirichlet boundary condition.
Let $G=G(x,y,k)$ be the resolvent kernel of the Dirichlet Laplacian in 
$D'$:
$$
(\Delta +k^2)G=-\d(x-y) \quad in \quad D', G|_{S}=0,
 \eqno{(3)}
$$
and $G$ satisfies the outgoing radiation condition.

If 
$$
(\Delta +k^2)w=0 \quad in \quad D', \quad w|_{S}=h,
 \eqno{(4)}
$$
and $w$ satisfies the radiation condition, then (\cite{R190})
one has
$$
w(x)=\int_S G_N(x,s)h(s)ds,\quad w=a(\a',k)e^{ikr}r^{-1} +o(r^{-1}), 
\quad as \quad r\to
\infty, \quad xr^{-1}=\a'.
\eqno{(5)}
$$
We write $a(\a')$ for $a(\a',k)$, and 
$$
a(\a'):=Bh:= \frac 1 {4\pi}\int_S u_N(s,-\a')h(s)ds,
 \eqno{(6)}$$
as follows from Ramm's lemma:

{\bf Lemma 1\,\,} (\cite{R190}, p.46) {\it One has:

$$
G(x,y,k)= g(r)u(y,-\a',k) +o(r^{-1}),
\quad as \quad r=|x|\to
\infty, \quad xr^{-1}=\a',
\eqno{(7)}
$$
where $u$ is the scattering solution (1)-(2).}

 One can write the scattering amplitude  as:
$$
A(\a',\alpha,k)=-\frac 1 {4\pi}\int_S u_N(s,-\a')e^{ik\alpha \cdot s}ds.
\eqno{(8)}
$$
The following claim is proved in \cite{Kir98}:

{\it Claim:}  $f:=e^{-ik \alpha' \cdot z}\in R(B)$ if and only if  
$z\in D$.

{\it Proof of the claim}. Our proof is based on the results in [7].

a) Let us assume  that $f=Bh$, i.e., $f\in R(B),$ and prove that $z\in D$.
Define $p(y):=g(y,z)-\psi(y)$, where $\psi(y):=\int_SG_N(s,y)h(s)ds$. 
The function $p(y)$ solves the Helmholtz equation (4) in the region 
$|y|>|z|$ and 
$p(y)=o(\frac 1 {|y|})$ as $|y|\to \infty$ because of (7) and of the
relation $Bh=f$. Therefore (see [7], p.25) $p=0$ in the region $|y|>|z|$.
Since $\psi$ is bounded in $D'$ and $g(y,z)\to \infty$ as $y\to z$, we get
a contradiction unless $z\in D$. Thus,  $f\in R(B)$ implies
$z\in D$.

b) Let us prove that $z\in D$ implies $f\in R(B)$. 
Define $\psi(y):=\int_SG_N(s,y)g(s,z)ds$, and $h:=g(s,z)$. Then, by
Green's formula, one
has $\psi(y)=g(y,z)$. Taking here $|y|\to \infty$, $\frac y
{|y|}=\alpha'$,
and using (7), one gets  $f=Bh$, so $f\in R(B)$. 
The claim is proved. $\Box$
 
Consider $B: L^2(S)\to L^2(S^2)$, and $A: L^2(S^2)\to L^2(S^2)$,
where $B$ is defined in (6) and $Aq:=\int_{S^2}A(\a',\alpha)q(\alpha) 
d\alpha.$ 
 
{\bf Theorem 1.} {\it The ranges $R(B)$ and $R(A)$ are dense in 
$L^2(S^2).$
}

{\bf Proof.} Recall that assumption (A) holds. It is sufficient to prove 
that  $N(B^*)=\{0\}$ and  $N(A^*)=\{0\}$.
Assume $0=B^*q=\int_{S^2}\overline {u_N(s,-\a')}qd\a'$, where the 
overline stands for complex conjugate. Taking complex conjugate and 
denoting $\overline {q}$ by $q$ again, one gets 
$0=\int_{S^2}u_N(s,-\a')qd\a'$. Define $w(x):=\int_{S^2} 
u(x,-\a')qd\a'$. Then $w=w_N=0$ on $S$, and $w$ solves equation (1) in 
$D'$. By the uniqueness of the solution to the Cauchy problem, $w=0$ in $D'$.
Let us derive from this that $q=0$. One has $w=w_0+V$, where 
$w_0:=\int_{S^2}e^{-ik\a' \cdot x}qd\a'$, and 
$V:=\int_{S^2}v(x,-\a',k)qd\a'$ satisfies the radiation condition.
Therefore, $w_0(x)=0$ in $D'$, as follows from Lemma 2 proved below.
By the unique continuation, $w_0(x)=0$ in
$\R^3$, and this implies $q=0$ by the injectivity of the Fourier 
transform. This proves the first statement of Theorem 1.
Its second statement is proved below. $\Box$ 

{\it Let us now prove Lemma 2, mentioned above.} 

We keep the notations used in the above proof.

{\bf Lemma 2.} If $w=w_0+V=0$ in $D'$, then $w_0=0$ in $D'$.

{\bf Proof.} The idea of the proof is simple: since $w_0$ does not satisfy
the radiation condition, and $V$ satisfies it, one concludes that 
$w_0=0$. Let us give the details. The key formula is (\cite{R190}, p.54):
$$\int_{S^2}e^{ik\alpha\cdot\b r}q(\b)d\b=\frac {2\pi i}k 
[\overline{\gamma} 
q(-\alpha) -\gamma q(\alpha)] +o(\frac 1 r), \quad r\to \infty,
\eqno{(9)}
$$ 
where $\gamma:=e^{ikr}/r$, and one assumes $q\in C^1(S^2)$.

If $r:=|x|\to \infty$, then, by Lemma 2, assuming $q\in C^1(S^2)$,
and using the relation $w=w_0+V=0$ in $D'$, 
one gets $q(\alpha)=0$ for all $\alpha\in S^2$. Thus, Lemma 2 is proved 
under the additional assumption $q\in C^1(S^2)$. If $q\in L^2(S^2)$,
then one uses a similar argument in a weak sense, i.e., with $x:=r\b$, one 
considers the inner product in $ L^2(S^2)$ of $w_0(r\b)$ and
a smooth test function $h\in C^\infty(S^2)$, and applies Lemma 2
to the function $\int_{S^2}e^{-ik\a' \cdot \b r}hd\b$.
Then, using arbitrariness of $h$, one concludes that $q=0$ as an
element of $L^2(S^2)$. Lemma 2 is proved. $\Box$. 

{\it Let us prove the second statement of Theorem 1.}

Assume now that $A^*q=0$. Taking complex conjugate,
 and using the reciprocity
relation: $A(\alpha,\b)=A(-\b,-\alpha)$, one gets an equation: 
$$\int_{S^2}A(\alpha,\b)hd\b=0,
\eqno{(10)}
$$ 
where $h=\overline {q(-\b)}$.
Define $w(x):=\int_{S^2}u(x,\b)hd\b$. Then $w=w_0+V$, where
$w_0:=\int_{S^2}e^{ik\b \cdot x}hd\b,$ and 
$V:=\int_{S^2}v(x,\b)hd\b$ satisfies the radiation condition.
Equation (10) implies that $V=o(\frac 1 r)$ as $r\to \infty$. Since
function $V$ solves equation (1) and $V=o(\frac 1 r)$, one concludes (see 
\cite{R190}, p.25), that $V=0$ in $D'$, so that $w=w_0$ in 
$D'$. Thus, $w_0|_S=w|_S=0$.
Since $w_0$ solves equation (1) in $D$ and  $w_0|_S=0$, one gets,
using Assumption (A), that $w_0=0$ in $D$. This and the unique 
continuation property imply $w_0=0$ in $\R^3$. Consequently, 
$h=0$, so $q=0$, as claimed.
Theorem 1 is proved. $\Box$ 

{\bf Remark 1.} In \cite{ck} the 2D inverse obstacle scattering 
problem is considered. It is proposed to solve the equation 
(1.9) in \cite{ck}:
$$\int_{S^1}A(\alpha,\b)\g d\b=e^{-ik\alpha \cdot z},
\eqno{(11)}
$$
where $A$ is the scattering amplitude at a fixed $k>0$, $S^1$ is the unit
circle, and $z$ is a point on $\R^2$. If $\g=\g(\b, z)$ is found, the 
boundary
$S$ of the obstacle is to be found by finding those $z$ for which
$||\g||:=||\g(\b,z)||_{L^2(S^1)}$ is maximal.
Assuming that $k^2$ is not a Dirichlet or Neumann eigenvalue
of the Laplacian in $D$, that $D$ is a smooth, bounded, simply
connected  domain, the authors state Theorem 2.1 \cite{ck}, p.386,
which says that for every $\epsilon>0$ there exists a function $\g\in 
L^2(S^1),$ such that 
$$
\lim_{z\to S}||\g(\b,z)||=\infty,
\eqno{(12)}
$$
and ( see \cite{ck}, p.386),
$$||\int_{S^1}A(\alpha,\b)\g d\b-e^{-ik\alpha \cdot z}||<\epsilon.
\eqno{(13)}
$$
 
 There are several questions concerning the proposed method.

First, equation (11), in general, is not solvable. The authors propose to 
solve it approximately, by a regularization method. The regularization 
method applies for stable solution of solvable
ill-posed equations (with exact or noisy data). If equation (11)  
is not solvable, it is not clear what numerical "solution"
one seeks by a regularization method.

Secondly, since the kernel of the integral operator in (11) is
smooth, one can always find, for any $z\in \R^2$, infinitely many $\g$ with 
arbitrary large $||\g||$, such that (13) holds.
 Therefore it is not clear how and why, using (12), one can find $S$ 
numerically by the proposed method.

{\bf Remark 2.} In [2], p.386, Theorem 2.1, it is claimed that 
for every $\epsilon>0$ and every $y_0\in D$ there exists a function $\g$ 
such that inequality (13) (which is (2.8) on p.386 of [2]) holds and $||\g||\to \infty$
as $y_0\to \partial D$. Such a $\g$ is used in [2] in a "simple method for 
solving inverse scattering problem". However, in fact there exist
infinitely many $\g$ such that inequality (13) holds and 
$||\g||\to \infty$, 
regardless of where $y_0$ is. Therefore it is not clear how one can use 
the method proposed in [2] for solving the inverse scattering problem with 
any degree of confidence in the result. 
 
{\bf Remark 3.} In \cite{b} it is mentioned that the methods
(called LSM-linear sampling methods) 
proposed in papers \cite{cc}, \cite{ck}, \cite{Kir98} 
produce numerically results which are inferior to these obtained
by linearized Born-type inversion. There is no guarantee of any accuracy 
in recovery of the obstacle by LSM. Therefore it is of interest to 
experiment numerically with other inversion methods. In  
\cite{R190}, p.94, (see also  \cite{R278},\cite{R172},\cite{R173})
a method (SFM-support function method) is proposed for recovery of 
strictly convex obstacles from the scattering amplitude.
This method allows one to recover the support function of the obstacle, 
and the boundary of the obstacle is obtained from this function 
explicitly. Error estimates of this method are obtained for the case when 
the data are noisy \cite{R190}, p.104. The method is asymptotically exact 
for large 
wavenumbers, but it works numerically even for $ka\sim 1$, as shown in 
\cite{GR}. For the Dirichlet, Neumann and Robin boundary conditions
this method allows one to recover the support function without a priori 
knowledge of the boundary condition. If the obstacle is not convex,
the method recovers the convex hull of the obstacle. Numerically
one can recover the obstacle, after its convex hull is found,
by using Modified Rayleigh conjecture method, introduced in \cite{R430},
or by a parameter-fitting method. 

In \cite{R325} there is a formula for finding an acoustically soft 
obstacle from the fixed-frequency scattering data. It is an open problem 
to develop an algorithm based on this formula.

A numerical implementation of the Linear Sampling Method (LSM), suggested 
in \cite{ck},
consists of solving a discretized version of 
$$
\int_{S^1}A(\alpha,\b)\g d\b=e^{-ik\alpha \cdot z},\eqno{(14)}
$$
where $A$ is the scattering amplitude at a fixed $k>0$, $S^1$ is the unit
circle, $\alpha\in S^1$, and $z$ is a point on $\R^2$. 

Let $F=\{A{\alpha_i,\b_j}\},\; i=1,...,N$, $j=1,...,N$ be the square matrix
formed by the measurements of the scattering amplitude for $N$ incoming,
and $N$ outgoing directions. then the discretized version of (14) is
$$
F\bg=\bff,\eqno{(15)}
$$
where the vector $\bff$ is formed by
$$
\bff_n=\frac{e^{i\frac \pi 4}}{\sqrt{8\pi k}}e^{-ik\alpha_n\cdot z},\quad
n=1,...,N,\eqno{(16)}
$$
see \cite{b} for details.

Denote the Singular Value Decomposition of the far field operator by 
$F=USV^H$. Let $s_n$ be the singular values of $F$, $\rho=U^H\bff$, and $\mu=V^H \bff$. Then the norm of
the sought function $g$ is given by
$$
\|\g\|^2=\sum^N_{n=1}\frac{|\rho_n|^2}{s_n^2}.    \eqno{(17)}
$$
A different LSM is suggested by A. Kirsch in \cite{Kir98}. In it one solves
$$
(F^*F)^{1/4}\bg=\bff  \eqno{(18)}
$$
instead of (15). The corresponding expression for the norm of
$\g$ is
$$
\|\g\|^2=\sum^N_{n=1}\frac{|\mu_n|^2}{s_n}. \eqno{(19)}
$$
A detailed numerical comparison of the two LSMs and the linearized
tomographic inverse scattering is given in \cite{b}. 

 The conclusions of \cite{b}, as well as of our own numerical
experiments are that the method of Kirsch (19) gives a better,
but a comparable identification, than (17). The identification
is significantly deteriorating if the scattering amplitude is available
only for a limited aperture, or if the data are corrupted by noise. Also, 
the points with the
{\it smallest} values of the $\|\g\|$ are the best in locating the
inclusion, and not the {\it largest} one, as required by the theory in \cite{Kir98}
and in \cite{ck}.
 In Figures 1 and 2 the 
implementation of the
Colton-Kirsch LSM (17) is denoted by $gnck$, and of the
Kirsch method (19) by $gnk$. The Figures show a contour plot of the logarithm
of the $\|\g\|$. The original
obstacle consisted of two circles
of radius $1.0$ centered at the points $(-d,\ 0.0)$ and $(d,\ 0.0)$. 
The results of the identification for $d=2.0$ are shown in Figure 1, and the results for $d=1.5$
are shown in Figure 2.
Note that the actual radius of the circles is $1.0$, but it cannot be seen from the
LSM identification. Also, one cannot determine the separation between the circles, nor their shapes.
Still, the methods are fast, they locate the obstacles, and do not require any knowledge of the
boundary conditions on the obstacle. The Support Function Method (\cite{GR}, \cite{R190}) 
showed a better identification for the convex parts of obstacles. Its generalization for unknown
boundary conditions is discussed in \cite{R463}. The LSM identification
was performed for the scattering amplitude of the obstacles computed
by the Boundary Integral Equations method, see \cite{coltonkress}.
No noise was added to the synthetic data. 
 In all the experiments we used $k=1.0$, and $N=60$.

\begin{figure}
\includegraphics*{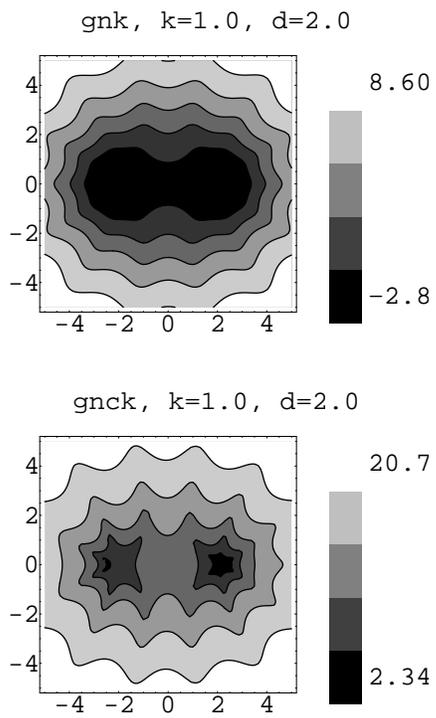} 
\caption{Identification of two circles of radius $1.0$ centered at 
$(-d,\ 0.0)$ and $(d,\ 0.0)$ for $d=2.0$.}
\end{figure}

\begin{figure}
\includegraphics*{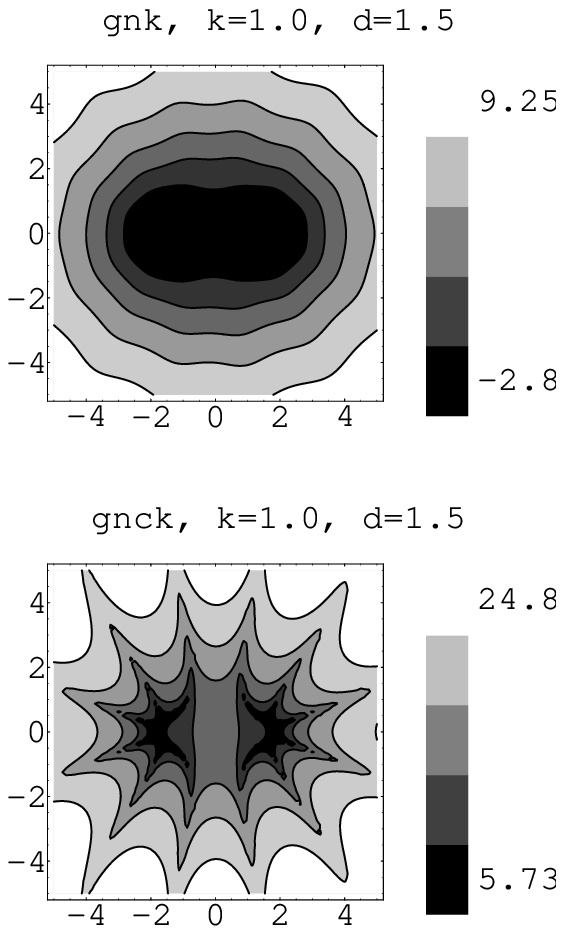} 
\caption{Identification of two circles of radius $1.0$ centered at 
$(-d,\ 0.0)$ and $(d,\ 0.0)$ for $d=1.5$.}
\end{figure}

\end{document}